\documentclass[a4paper,10pt,twocolumn]{article} 
\usepackage{bbm,amssymb,amsmath,amsbsy,times,pifont,pxfonts}
\usepackage{graphicx}
\usepackage{subfigure}
\usepackage{fancyhdr,makeidx}
\usepackage{color}

\newcommand{\N}{\cal N}

\newcommand{\id}{\mathbbm{1}}

\newcommand{\tr}{{\rm Tr}\,}
\renewcommand{\det}{{\rm Det}\,}
\newcommand{\gr}[1]{\boldsymbol{#1}}
\newcommand{\be}{\begin{equation}}
\newcommand{\ee}{\end{equation}}
\newcommand{\bea}{\begin{eqnarray}}
\newcommand{\eea}{\end{eqnarray}}

\newcommand{\sig}{\gr{\sigma}}

\newcommand{\eq}[1]{Eq.~(\ref{#1})}

\newcommand{\proofend}{\hfill\fbox\\\medskip }

\newcommand{\figbox}[1]{%
  \fbox{%
    \vbox to 1.4in{%
    \vfil
    \hbox to 1in{%
      \hfil
      #1%
      \hfil}%
    \vfil}}}

\title{
\normalsize \bf \textsf{\large \hspace*{-.2cm}\mbox{Would one rather store squeezing or entanglement in continuous variable quantum memories?}}}

\author{\normalsize Hulya Yadsan-Appleby and Alessio Serafini}

\date{\small \it Department of Physics \& Astronomy\\ 
\small \it University College London\\
\small \it Gower Street, London WC1E 6BT}

\begin{document}

\maketitle

\subsection*{\normalsize \bf\textsf{Abstract}}
{\em Given two quantum memories for continuous variables ({\em e.g.}, the collective pseudo-spin of
two atomic ensembles) 
and the possibility to perform passive optical operations (typically beam-splitters) 
on the optical modes before or after the storage, two possible scenarios arise resulting 
in generally different degrees of final entanglement. 
Namely, one could either store an entangled state and retrieve 
it directly from the memory, or rather store two separate single-mode squeezed states and then
combine them with a beam-splitter to generate the final entangled state.
In this paper, we address the question of which of these two options yields the higher entanglement.
By adopting a well established description of QND feedback memories, and 
a simple but realistic noise model, 
we analytically determine the optimal choice for several regions 
of noise parameters and quantify the advantage it entails, 
not only in terms of final entanglement but also in terms of the 
capability of the final state to act as a shared resource for quantum teleportation.
We will see that, for `ideal' or `nearly ideal' memories, 
the more efficient of the two options is the one 
that better protects the quadrature subject to the largest noise in the memory 
(by increasing it and making it more robust).}

\subsection*{}
Quantum memories are set to be a crucial component of future quantum computers. 
Even in the short and medium term, the development of effective quantum memories 
would pave the way for the implementation of a variety of quantum information protocols. 
For instance, in quantum communication -- where distributing quantum correlations over long 
distances is a primary, and difficult, requirement -- quantum repeaters assisted 
by good enough quantum memories could solve the problem of entanglement distribution
in a not so distant future. 
Hence, considerable efforts are currently being devoted to improving the 
performance of quantum memories, in terms of both reliability and storage times \cite{review}. 

Central among the controllable degrees of freedom that are benefiting from such technological developments 
are the so-called quantum continuous variables, as one customarily refers to the 
degrees of freedom described by pairs of canonical operators, 
like second quantized electromagnetic fields or motional degrees of freedom of massive particles. 
Continuous variables, mostly in their optical form, are very well suited for quantum communication 
and key distribution  
\cite{cvrmp,cvcrypto}, 
due to the ease with which they can be distributed and to the comparative reluctance to interact 
with their environment, and hence to undergo decoherence.
However, the very same reluctance that makes them so resilient in the face of environmental noise 
also implies that establishing quantum correlations 
between optical continuous variables, as necessary to exploit the advantages offered by 
quantum communication, is very difficult, 
and that only a very limited amount of entanglement can be generated in practice. 
A standard scheme to obtain continuous variable entanglement consists in mixing 
two single-mode squeezed states -- with properly oriented optical phases -- into a 50:50 
beam-splitter. More generally, continuous variable entanglement always requires some 
degree of squeezing: the statistical variance of some (composite) 
degree of freedom of the system has to be below one unit of vacuum noise
for entanglement to exist \cite{passive}. 
Continuous variable squeezing and entanglement are therefore  
closely related resources 
in the sense that, given the former, the latter can be generated by 
``passive'' optical operations, like beam-splitting, which are 
much less demanding than squeezing to realise in the laboratory.

Besides the production of quantum entanglement, the other, already mentioned, basic ingredient for 
the implementation of quantum communication is the capacity 
to store states reliably in quantum memories.
Over the last ten years, a number of candidate strategies to store flying 
optical continuous variables into static degrees of freedom, typically utilising atomic ensembles, 
has been put forward and developed.
Notwithstanding the potential advantages offered by other schemes 
-- like, notably, EIT-based approaches \cite{eit1,eit2} -- 
the most promising, and currently most effective, among these strategies are 
based on quantum non demolition (QND) feedback.
In such schemes the light degrees of freedom are mapped into, and then successively retrieved from, 
the collective pseudo-spin of an atomic cloud \cite{rmp}. This approach has allowed for the 
demonstration of the proper, `quantum' storage of coherent states \cite{expmemo} and for the 
teleportation of quantum information between light and matter \cite{exptele}. 

\begin{figure}[t!]
\begin{center}
\subfigure[\label{casea}]{\figbox{\includegraphics[scale=0.6]{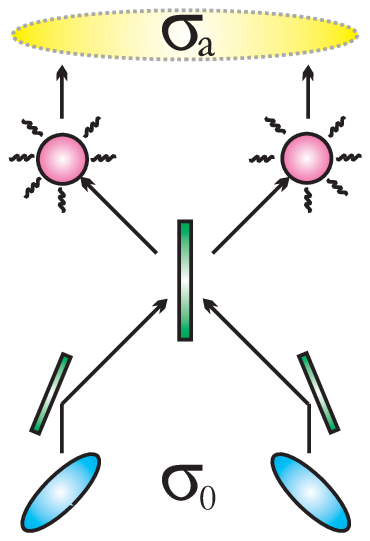}}}
\hspace*{1cm}
\subfigure[\label{caseb}]{\figbox{\includegraphics[scale=0.6]{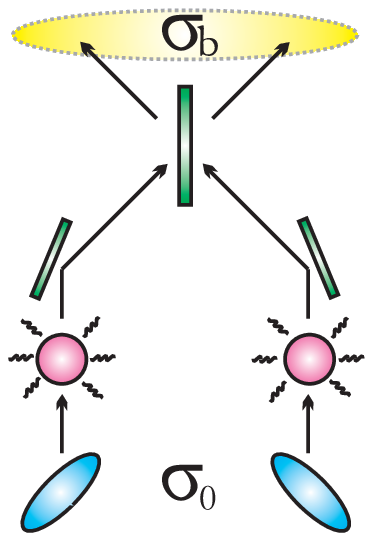}}}
\caption{\small The two cases being compared: in (a), the state is stored in the memory cells {\em after} the 
beam splitter has mixed and entangled the single-mode squeezed states; 
in (b), the state is stored {\em before}
the entangling beam-splitting action. We assume, conceding something to idealisation, 
that all the noise is imputable to the storage and retrieval process. 
In practice, such a noise would certainly dominate over the mixing's imperfections.\label{cases}}
\end{center}
\end{figure}

Very recently, the successful quantum storage of squeezing and entanglement 
in these memories has been reported as well \cite{squeezememo}. 
The possibility of storing entanglement and squeezing poses a 
relevant theoretical question: given some precious single-mode squeezing, which has to be generated 
in costly and delicate parametric down conversion processes, is it better to store it as it is
in a quantum memory, and then use it to produce entanglement after retrieval from the 
memory, or would one rather first use it to create entanglement and then map 
and retrieve the global entangled state? 
In other words, is one better off by storing squeezing or entanglement 
in a continuous variable quantum memory?
This inquiry is dedicated to such a question.
We will specialise to QND feedback memories, and carry out a 
comparative study between two cases, labeled with $a$ and $b$ respectively,
where the memory will act after or before an entangling 50:50 beam-splitter (see Fig.~\ref{cases}).
The figure of merit we will adopt is the final entanglement obtained in the two cases,  
{\em i.e.}\ the entanglement that, in such
scenarios, would be available ``on demand'' after the memory's operation.
We will also extend our study to a directly operational figure of merit: the capability 
of the final state to act as a shared channel for the quantum teleportation
of coherent states.

\subsection*{\normalsize \bf \textsf{Notation and preliminaries}} 
Since all the states involved in this problem are, with very good approximation, Gaussian states, 
we can take advantage of the Gaussian formalism \cite{jensin} in our analysis, 
whereby quantum states 
will be entirely characterised by their covariance matrices, whose entries represent 
all the second 
statistical moments of the canonical operators (first moments, being independent from the correlations, 
are irrelevant to our discussion and will henceforth be neglected). 

If the two-mode quantum system in question is described by the vector of canonical operators 
${\bf \hat{r}} = \{\hat{q}_1,\hat{p}_1,\hat{q}_2,\hat{p}_2\}$, 
the covariance matrix (CM) $\sig$ of the state $\varrho$ is defined as 
$\sigma_{jk} = \tr[\varrho\{\hat{r}_j,\hat{r}_k\}] - 2 \tr[\varrho \hat{r}_j]\tr[\varrho \hat{r}_k]$.
In all that follows, we shall adopt natural units ($\hbar=1$) and 
report all the variances in units of vacuum noise (such that the vacuum state has CM equal to 
the identity matrix).

In this notation, the action of the memory can be described 
as a Gaussian quantum channel, acting on the covariance matrix $\sig$ as \cite{Gchannels}
$$
\sig \rightarrow X\sig X^{\sf T} + Y \; ,
$$
with $X={\rm diag}(\xi_1,\xi_1,\xi_2,\xi_2)$ and 
$Y={\rm diag}(y_{q1},y_{p1},y_{q2},y_{p2})$  
(that is, the original state is recovered up to some multiplicative losses and additive noise 
on the system's canonical operators). 
The relationships $\xi_i^2\ge (1-\sqrt{y_{qi}y_{pi}})$ for $i=1,2$ guarantee that 
the channel is a physical map \cite{Gchannels}.
Notice that this description accounts for all the noise observed in experiments, 
including losses at the input and output of the memory cells, 
uncertainties in the initial atomic collective pseudo-spin along the direction 
which is not addressed by the feedback loop, 
and additional noise of spurious nature (atomic decoherence, unsuppressed noise due to 
imperfections in the feedback loop, further technical noise, see  
\cite{squeezememo}).
Later on, we will see how the matrix $Y$ relates to practical parameters,
and consider realistic values of such parameters. For the moment being, let us just 
mention that in an `ideal' implementation of the memory, where the only residual noise 
is due to the initial uncertainty in the atomic pseudo-spin, one would have $\xi_1=\xi_2=1$ 
and $y_{p1}=y_{p2}=0$ \cite{rmp}.
This operating regime will be referred to as `ideal memories' in the following.
Since each atomic cloud will be generally different from another, 
$\xi_1$, $y_{q1}$ and $y_{p1}$ will generally differ from $\xi_2$, $y_{q2}$ and $y_{p2}$.

The two possibilities we are about to compare can be described by two different  
final quantum states, with different covariance matrices: 
$\varrho_a$, with CM $\sig_a$, where the noisy channel describing the memory 
is applied {\em after} the beam-splitter, 
corresponding to storing entanglement [Fig.~\ref{casea}], and $\varrho_b$, with CM $\sig_b$, 
where the noisy channel is applied 
{\em before} the beam-splitter, corresponding to the storage of squeezing [Fig.~\ref{caseb}].
One has
\bea
\sig_{a} &=& X R \sig_0 R^{\sf T} X^{\sf T} + Y \; , \label{siga} \\
\sig_{b} &=& R X\sig_0X^{\sf T} R^{\sf T} + RY R^{\sf T} \; , \label{sigb}
\eea
where $R={\rm e}^{J \pi/4}$ describes the action of the 50:50 beam-splitter
in phase space, in terms of the $4\times 4$ anti-symmetric generator $J$ with entries 
$J_{jk} = \delta_{j+2,k}-\delta_{j,k+2}$ for $1\le j,k\le 4$.
Note that  
if the noises in the two memories were perfectly identical then 
$[R,X]=[R,Y]=0$ and $\sig_a=\sig_b$, so that there would be no difference between 
the two cases.

For simplicity, and to fix ideas, we shall assume an initial CM given by:
$
\sig_0 = {\rm diag}
(s N_1,N_1/s,N_2/s,N_2 s)
$,
with $s\ge 1$, $N_1\ge 1$ and $N_2\ge 1$.
For $N_1=N_2=1$, this CM describes two pure single-mode squeezed states with optical 
phases chosen so as to optimise the production of entanglement by a 50:50 
beam-splitter \cite{passive}. 
Two different squeezing parameters could be chosen for the 
two modes but, in this context, it seems reasonable to endow the hypothetical experimenter 
with a fixed squeezing capability (different squeezing parameters can be treated 
with our approach, but would just complicate the study without introducing any 
substantial conceptual difference). Finally, the parameters $N_1$ and $N_2$ describe 
the thermal broadening squeezed states are likely to undergo in practice. 

The entanglement of the quantum states $\varrho_a$ and $\varrho_b$ 
can be characterised 
by evaluating their logarithmic negativities $E_{\N}(\varrho_a)$ and $E_{\N}(\varrho_b)$, 
an entanglement monotone which represents 
an upper bound -- in `ebits' -- to the distillable entanglement,
{\em i.e.}\ to the maximal number of Bell pairs per copy of the entangled state 
which can be obtained by local operations and classical communications alone \cite{nega}.
The logarithmic negativity is, for Gaussian states such as these, 
an increasing function of the
smallest partially transposed symplectic eigenvalue $\tilde{\nu}$, which is in turn a function of 
two partially transposed symplectic invariants $\det\sig$ and $\tilde{\Delta}(\sig)$ \cite{serafozzi}. 
If one expresses the covariance matrices in terms of $2\times2$ blocks, as in 
\be
\sig_{i} = \left(\begin{array}{cc}
{\gr \alpha}_i & {\gr \gamma}_i\\
{\gr \gamma}_I & {\gr \beta}_i
\end{array}\right) \quad {\rm for} \quad i=a,b \; , \label{2x2}
\ee
the partially transposed symplectic invariants are $\det{\sig}_i$ and 
$\tilde\Delta(\sig_i) = \det{\gr \alpha}_i + \det{\gr \alpha}_i-2\det{\gr \gamma}_i$ for $i=a,b$, 
and determine the smallest symplectic eigenvalue as follows
\be
2\tilde{\nu}_{i}^2=\tilde{\Delta}(\sig_i)-\sqrt{\tilde{\Delta}^2(\sig_i)-4\det\sig_{i}} 
\quad {\rm for}\quad i=a,b \, . \label{numin}
\ee
Finally, the logarithmic negativity is given, in ${\rm ebits}$, by 
\be
E_{\N}(\varrho_i) = \max\left[0,-\log_2 \tilde{\nu}_i\right] \quad {\rm for} \quad i=a,b \, . \label{logneg}
\ee
Note that for a state to be entangled one must have $\tilde{\nu}<1$.

\subsection*{\normalsize \bf \textsf{Ideal memories}}

In the case $\xi_1=\xi_2$ and $y_{p1}=y_{p2}=0$, we could derive a sharp analytical criterion to 
discriminate between the storage of squeezing and entanglement:

{\em Given the notation of the previous section and assuming 
$y_{p1}=y_{p2}=0$, $\xi_{1}=\xi_2$ and
\be
\frac{1}{s^2}\le \frac{N_2}{N_1} \le s^2 \; , \label{assume}
\ee
one has
\bea
y_{q2}\ge y_{q1} &\Leftrightarrow& E_{\N}(\varrho_{a}) \ge E_{\N}(\varrho_b) \; . \label{finda} 
\eea}
That is, given this configuration of optical phases (fixed by the condition $s\ge 1$), 
storing entanglement is advantageous over storing single-mode squeezing if the noise 
acting on the second quadrature is larger than the noise acting on the first quadrature, 
and viceversa. The proof of the statement above can be found in appendix.
Notice that the assumption (\ref{assume}) on the input state
is very mild in that, even for a (very reasonable) 
squeezing parameter $s=4$ (corresponding 
to $6$ ${\rm dB}$ of squeezing), 
the assumption is violated if the input thermal noise affecting one mode is 
more than $16$ times larger than the noise acting on the other mode. 
This condition will thus be met in most practical instances, which renders the criterion 
effectively independent from the parameters 
of the input state, as desirable. 

Besides providing us with an analytical criterion valid in a relevant case, the finding above 
also sheds light on the physical origin of the advantage granted by one or the other strategy.
{\em Given this configuration of optical phases}, 
storing entanglement ({\rm i.e.}\ applying the entangling beam-splitter 
before the retrieval from the memory) is advantageous if the noise acting on the second 
mode is larger than the noise acting on the first one: this is because, since $s>1$, 
storing squeezing would mean that the larger noise ($y_{q2}$) 
would act on the (more delicate) squeezed quadrature $\hat{q}_{2}$. 
More generally, {\em the optimal storage is the one whereby the variance of the 
noisiest quadrature is the larger before the storage takes place, 
and hence is the more robust in the face of the noise}.
Hence, our findings point to a way to improve the storage of 
continuous variable entanglement, provided 
that the additive noise that characterises each single-mode memory is known {\em a priori}. 
This would certainly be possible in practice, since each memory could be calibrated before use.
However, it should be stressed that the relative advantage of storing two-mode 
entangled states or single-mode squeezing can only be determined if the noises associated to the two memory cells are known.

Let us now briefly consider the quantitative advantage granted by the optimal choice by considering values taken 
from practical experiments. Under ideal conditions, one would have $y_{q1}=(1-1/Z_1^2)\Delta_{At1}$ 
and $y_{q2}=(1-1/Z_2^2)\Delta_{At2}$. Here, $Z_1$ and $Z_2$ are parameters 
which depend only on the optical detuning of the swap interaction 
and take the value $\sqrt{6.4}$ in our experiment of reference \cite{squeezememo}, 
whereas $\Delta_{At1}$ and 
$\Delta_{At2}$ are the initial variances of one of the 
quadratures of the collective atomic pseudo-spin in the two memory cells. 
The atomic clouds are initialised in spin-squeezed states in order to reduce
$\Delta_{At1}$ and $\Delta_{At2}$. 
Although in principle the spin-squeezing of the atomic ensembles could be pushed further,
in practical instances these quadratures still take values around $0.8$ (in units of vacuum noise). 
It is reasonable to assume that the main difference between the noise added by the two memories would be due to differences between $\Delta_{At1}$ and $\Delta_{At2}$,
which are far less controllable than $Z_1$ and $Z_2$.
Hence, assuming $Z_1^2=Z_2^2=6.4$, $s=4$, 
$N_1=N_2=1$, $\Delta_{At1}=0.6$ and $\Delta_{At2}=1$,
one would have $E_{\N}(\varrho_b)=1.06$ ebits and $E_{\N}(\varrho_a)=0.94$ ebits: 
in this case, storing single-mode squeezing rather than an entangled state 
would protect $0.12$ ebits of entanglement. 
As a rough rule of thumb, confirmed also for noisy input states ({\em i.e.}, for $N_1$ and $N_2$ 
not equal $1$), a difference of $0.25$ vacuum units in the additive noises of the two cells is reflected 
by a difference of $0.1$ ebits in the entanglement of $\varrho_a$ and $\varrho_b$.


\begin{figure}[t!]
\begin{center}
\subfigure[\label{comparea}]{\includegraphics[scale=0.46]{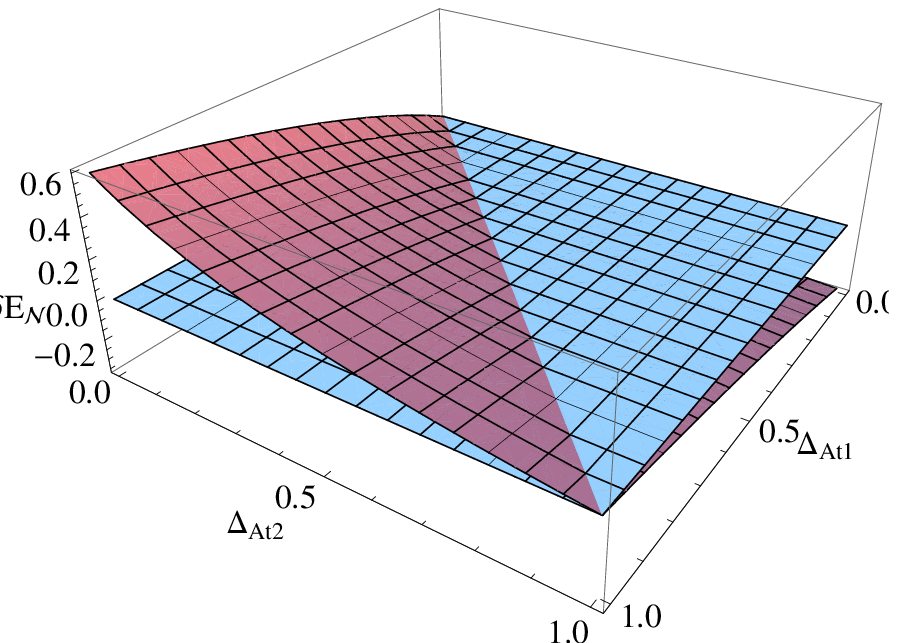}}
\hspace*{.2cm}
\subfigure[\label{compareteleb}]{\includegraphics[scale=0.46]{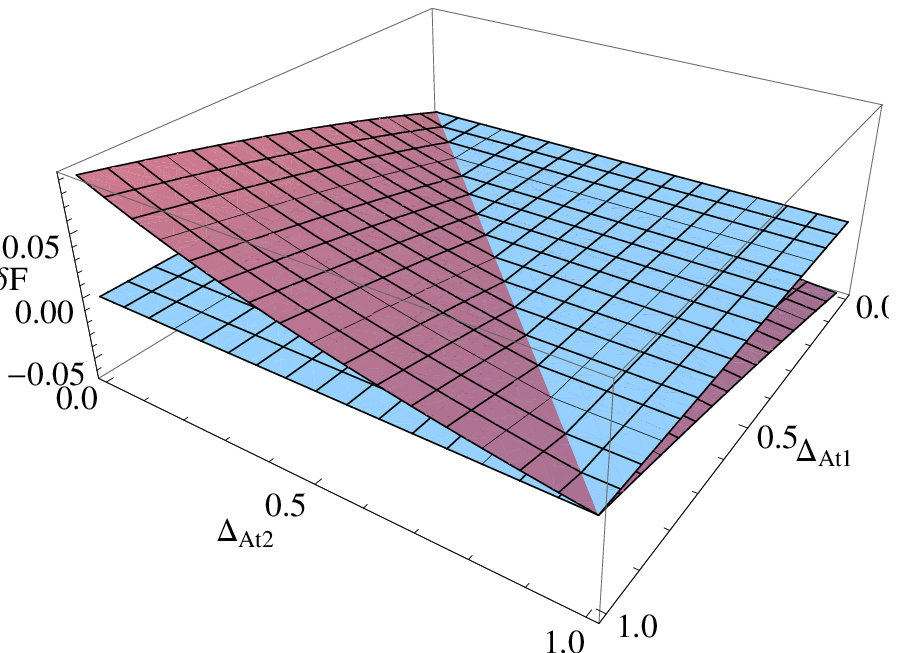}}
\end{center}
\caption{\small Difference between the logarithmic negativities $\delta E_{\N}=
E_{\N}(\varrho_b)-E_{\N}(\varrho_a)$ (a) and achievable teleportation fidelities 
$\delta F = F_a-F_b$ (b)
for $N_1=1.4$, $N_2=1.2$, $s=8$, $Z_1^2=Z_2^2=6.4$, 
$G_1=G_2=0.95$, $\Delta_{q1}=
\Delta_{q2}=0.1$ and $\Delta_{p1}=\Delta_{p2}=0.3$, 
as a function of $\Delta_{At1}$ and $\Delta_{At2}$.
The planes $\delta E_{\N}=0$ and $\delta F=0$ are also plotted,  
showing that  
$\varrho_b$ is more entangled when $\Delta_{At1}>\Delta_{At2}$. \label{compare}}
\end{figure}
\subsection*{\bf \normalsize\textsf{Noisy memories}}

In practice, the functioning of the quantum memories we are dealing with 
is not only limited by the initial variance of the atomic spin cloud, but also 
subject to input losses (as the light impinging on the memory cell 
is partially reflected) 
and to spurious additive noise, which has been estimated in recent 
experiments (see the Supplementary Material of Ref.~\cite{squeezememo}).
For $y_{p1}$ and $y_{p2}$ non-vanishing and $\xi_1\neq\xi_2$, no simple general criterion 
to determine which of $\varrho_a$ or $\varrho_b$ is more entangled could be 
derived. However, a systematic analytical comparison between the two cases was carried out.
Such a comparison shows that $\delta_{q}\le 0$ and 
$\delta_p \ge0$ $\Leftrightarrow$ $\delta E_{\N}\ge0$, 
where the new parameters are defined as $\delta_{q}\coloneqq y_{q2}-y_{q1}$,
$\delta_{p}\coloneqq y_{p2}-y_{p1}$ and $\delta E_{\N}\coloneqq 
E_{\N}(\varrho_{b})-E_{\N}(\varrho_a)$. 
If these assumptions on the additive noise are not satisfied, 
the optimal choice cannot be established in general, as the 
contribution of the noise acting on the $\hat{q}$ 
quadratures counteracts the noise on the $\hat{p}$ quadratures. 
The primary heuristic way to discriminate between the two cases is then simply to compare 
the differences in additive noises: thus one observes that  
if $\delta_p\ge \delta_q\ge 0$ or $\delta_q \le \delta_p\le 0$ then $\delta E_{\N}\ge 0$, 
for a wide range of parameters. 
The converse of this criterion, where $\delta_q \ge \delta_p \ge0$ and 
$\delta {E_{\N}}\le 0$, is however very often violated: 
{\em regardless of the chosen optical phase} ({\em i.e.}, even for $s<1$), the 
optimality of the storage of squeezing turns out to be more stable with respect to  
oscillations in the additive noise. 
Accordingly, in all observed instances we found that, 
for $y_{p1}=y_{q1}$ and $y_{q2}=y_{p2}$, one has
$\delta E_{\N} \ge 0$: the storage of squeezing is more robust 
in the presence of `phase-insensisitive' noise in the memories.

\begin{figure}[t!]
\begin{center}
\subfigure[\label{compga}]{\includegraphics[scale=0.46]{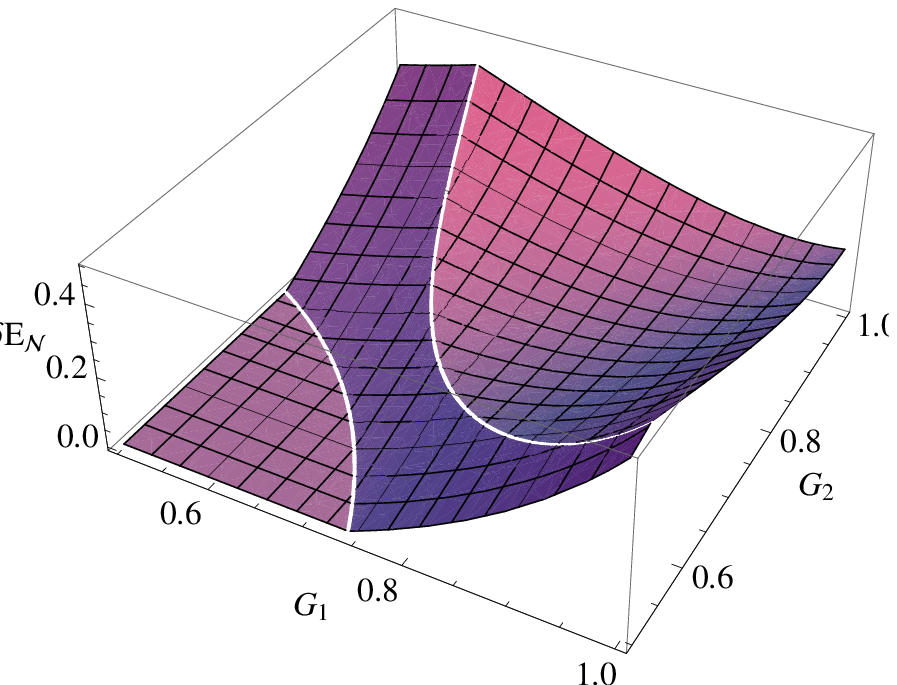}}
\hspace*{.2cm}
\subfigure[\label{compgb}]{\includegraphics[scale=0.46]{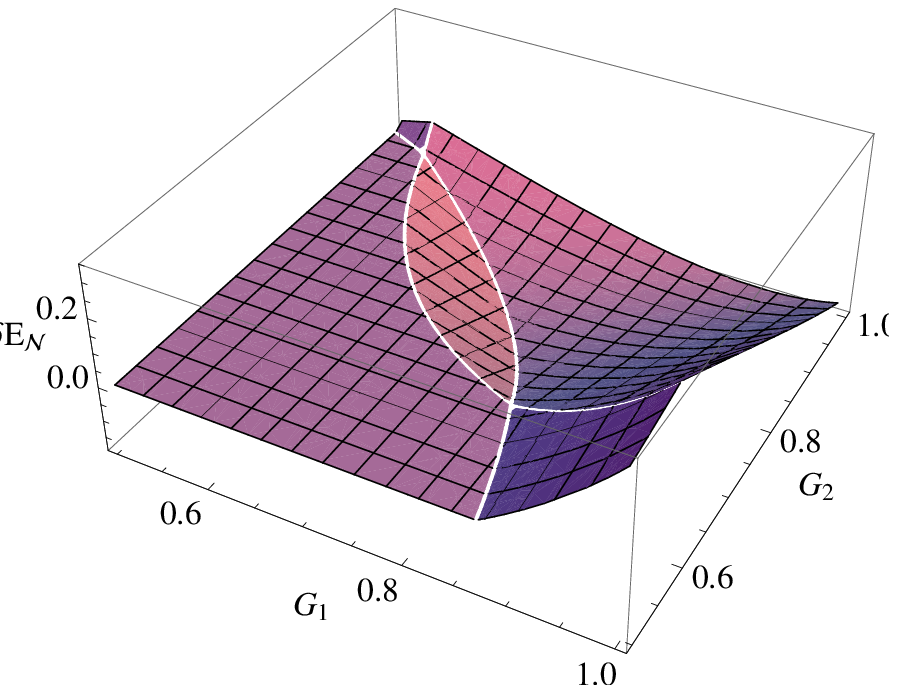}}
\caption{\small Dependence of the difference in the final entanglement on the loss factors. 
In (a), $\delta E_{\N}=E_{\N}(\varrho_b)-E_{\N}(\varrho_a)$ is plotted 
for $N_1$, $N_2=1$, $s=8$, $Z_1^2=Z_2^2=6.4$, 
$\Delta_{At1}=0.8$, $\Delta_{At2}=0.4$, $\Delta_{q1}=
\Delta_{q2}=0.1$ and $\Delta_{p1}=\Delta_{p2}=0.3$, 
as a function of $G_1$ and $G_2$; in this regime, the difference is positive for $G_1=G_2=1$, 
and stays so even varying the loss factors. In (b), the same values as in (a) are taken, except for 
$\Delta_{At1}=0.4$ and $\Delta_{At2}=0.8$; here the difference is negative with no losses, 
but can become positive by varying the loss factors. \label{compg}}
\end{center}
\end{figure}

In order to apply our study to a realistic experimental setting, 
let us now borrow the notation of reference \cite{squeezememo}, 
whereby $\xi_1=G_1$, $\xi_2=G_2$, and 
the entries of the $Y$ matrix would be 
$y_{q1} = (1-1/Z_1^2)\Delta_{At1}+(1-1/G_1^2)+\Delta_{q1}$, 
$y_{p1} = (1-1/G_1^2)+\Delta_{p1}$,
$y_{q2} = (1-1/Z_2^2)\Delta_{At2}+(1-1/G_2^2)+\Delta_{q2}$ and 
$y_{p2} = (1-1/G_2^2)+\Delta_{p2}$, 
where $G_1$ and $G_2$ quantify the losses, $\Delta_{q1}$, $\Delta_{p1}$, $\Delta_{q2}$ 
and $\Delta_{p2}$ represent the additional noise, of various origin (accounting for the decoherence of the 
atoms during the storage and for other imperfections), while $Z_1$, $Z_2$, 
$\Delta_{At1}$ and $\Delta_{At2}$ were defined in the previous section.
A typical comparison between case $a$ and $b$ in realistic conditions 
is shown in Fig.~\ref{comparea}. 
For the set of parameters $N_1=N_2=1$, $s=5$, $Z_1=Z_2=6.4$, $G_1=G_2=0.85$, $\Delta_{q1}=
\Delta_{q2}=0.2$, $\Delta_{p1}=\Delta_{p2}=0.4$, $\Delta_{At1}=0.9$ and $\Delta_{At2}=0.6$, 
which is well within current experimental reach, 
the state $\varrho_{a}$ is not entangled whereas the state $\varrho_b$ is entangled. 
In this instance, storing squeezing rather than entanglement into the 
memories would make the difference between an entangled or separable final state.

The effect of different loss factors in the two memory cells has also been investigated. 
If $G_1\neq G_2$, one can actually discriminate on more general grounds between the storage of squeezing or entanglement: an extensive quantitative analysis indicates that 
if the other parameters are such that, for $G_1=G_2=1$, $E_{\N}(\varrho_b)\ge E_{\N}(\varrho_a)$, 
then the same inequality holds for different $G_1$ and $G_2$ as well (see Fig.~\ref{compga}). 
Instead, as apparent in Fig.~\ref{compgb}, if the other parameters are such that 
$E_{\N}(\varrho_a)\ge E_{\N}(\varrho_b)$, this inequality may be reversed by varying the
loss factors. In this sense, storing squeezing would offer an additional guarantee 
in practical instances where the loss factors may vary widely between different memory cells.

\subsection*{\normalsize\bf\textsf{Teleportation fidelities}}

To put our results into a clear operational context, we intend now to adopt a different 
figure of merit in lieu of the logarithmic negativity: 
we will compare the storage of squeezing with the storage of entanglement 
in terms of how well the final states obtained allow one to perform the quantum teleportation of 
coherent states. The quality of the latter process can be estimated by the teleportation fidelities 
$F_a$ and $F_b$, defined as the overlap between the initial state and the teleported state 
when using, respectively, $\varrho_a$ and $\varrho_b$ as a shared entangled resource between 
sender and receiver \cite{cvrmp}. Although they are dependent on the entanglement of $\varrho_a$ 
and $\varrho_b$, $F_a$ and $F_b$ cannot be expressed in terms of $\tilde{\nu}_a$ and 
$\tilde{\nu}_b$ for general Gaussian states (see later), 
so our previous analysis does not, strictly speaking, imply any precise 
result concerning teleportation fidelities. 
However, $F_a$ and $F_b$ can be treated analytically in our setting. 

In terms of the $2\times2$ submatrices of \eq{2x2}, the optimal teleportation fidelity that can 
be achieved for input coherent states is given by
\be
F_{i} = \frac{2}{\sqrt{\det{(2\id + {\gr \alpha}_i + {\gr \beta}_i-2 \sigma_z{\gr \gamma}_i)}}} \, ,
\quad {\rm for}\quad i=a,b \, ,
\ee
where $\id$ is the two-dimensional identity matrix and $\sigma_z$ is the Pauli $z$ matrix \cite{piro}.
Some basic algebra leads to the following relationships for the fidelities:
${1}/{F^2_a} =
[1+({y_{p1}+y_{p2}})/{2}+\frac{N_1}{s}\left({\xi_1+\xi_2}\right)^2/4 + 
N_2 s \left({\xi_1-\xi_2}\right)^2/4 ] 
[1+{(y_{q1}+y_{q2})}/{2}+{N_2}/{s}\left({\xi_1+\xi_2}\right)^2/4 + 
N_1 s \left({\xi_1-\xi_2}\right)^2/4]$ and 
${1}/{F^2_b} = [1+y_{p1}+\frac{N_1}{s}\xi_1^2][1+y_{q2}+\frac{N_2}{s}\xi_2^2]$.
Upon immediate inspection, 
these equations allow one to retrieve the analogous of criterion (\ref{finda}) 
for the teleportation fidelities. For $y_{p1}=y_{p2}=0$ (`ideal' memories) 
and $\xi_1=\xi_2$, one has
\be
F_a\ge F_b \quad \Leftrightarrow \quad y_{q2}\ge y_{q1} \; . \label{findtele}
\ee
If none of the diagonal entries of $Y$ vanish or if the loss factors $\xi_1$ and $\xi_2$ are different, 
the situation is slightly more involved, 
since the comparison between $F_a$ and $F_b$ depends in general also on the parameters of the 
initial state $N_1$, $N_2$ and $s$. 

\begin{figure}[t!]
\begin{center}
\subfigure[\label{compgtelea}]{\includegraphics[scale=0.46]{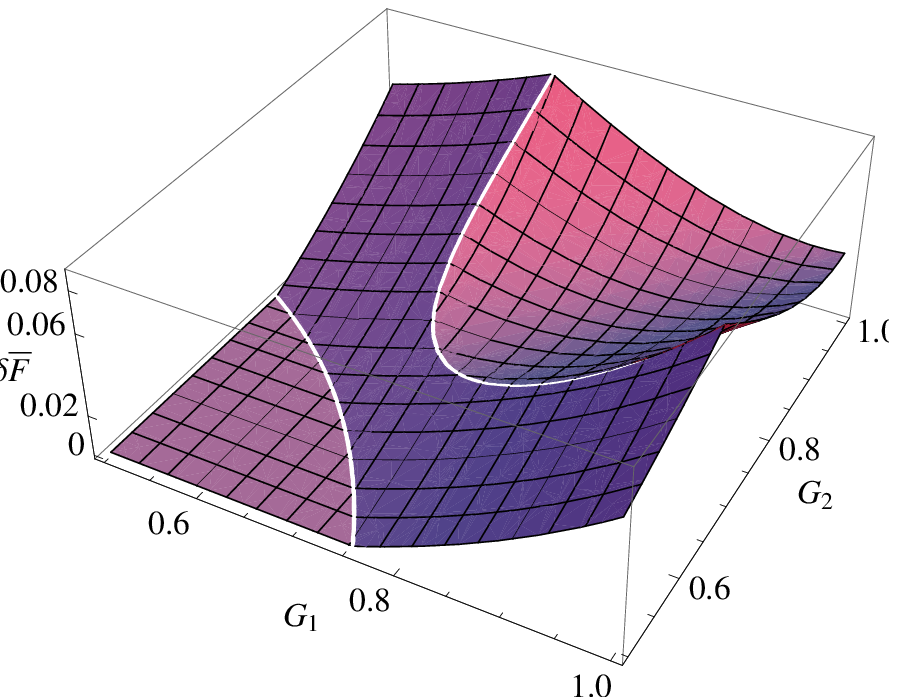}}
\hspace*{0.2cm}
\subfigure[\label{compgteleb}]{\includegraphics[scale=0.46]{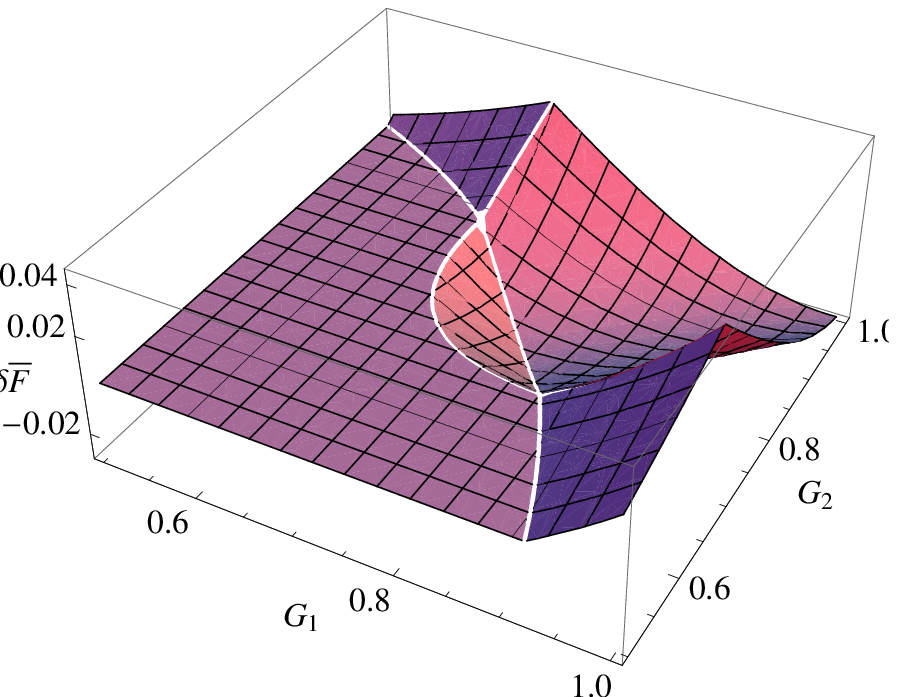}}
\caption{\small Dependence of the difference in teleportation fidelities 
$\delta \bar{F}=\max[F_b,1/2]-\max[F_a,1/2]$
on the loss factors. 
In (a), $\delta \bar{F}$ is plotted 
for $N_1$, $N_2=1$, $s=8$, $Z_1^2=Z_2^2=6.4$, 
$\Delta_{At1}=0.8$, $\Delta_{At2}=0.4$, $\Delta_{q1}=
\Delta_{q2}=0.1$ and $\Delta_{p1}=\Delta_{p2}=0.3$, 
as a function of $G_1$ and $G_2$. In (b), the same values as in (a) are taken, except for 
$\Delta_{At1}=0.4$ and $\Delta_{At2}=0.8$. Notice that the positive region in plot (b)
does not coincide exactly with that in plot \ref{compgb}. \label{compgtele}}
\end{center}
\end{figure}

However, as apparent when comparing Fig.~\ref{compareteleb} to Fig.~\ref{comparea} and 
Fig.~\ref{compgtele} to Fig.~\ref{compg}, 
the qualitative behaviour of the teleportation fidelities $F_a$ and $F_b$ reflects 
very closely that of 
the entanglement measures $E_{\N}(\varrho_a)$ and $E_{\N}(\varrho_b)$,
thus strengthening our previous analysis on operational grounds. 
Note that in Fig.~\ref{compgtele} we report the difference between the quantities 
$\max[F_i,1/2]$ for $i=1,2$, because $1/2$ is the threshold value that can be achieved by a `classical' 
measure-and-prepare strategy, without the help of state $\varrho$: this quantity better quantifies the 
advantage provided by the final states. 

Notice that the region of positive values in plot \ref{compgteleb} 
is different from the region of positive values in plot \ref{compgb}: {\em i.e.,} 
there are parameters' sets such that $F_b\ge F_a$  
but $E_{\N}(\varrho_a)\ge E_{\N}(\varrho_b)$. This fact leads to an interesting theoretical side remark: 
the teleportation fidelity of a two-mode Gaussian state 
cannot be related directly to the logarithmic negativity, 
not even for two-mode Gaussian states (at least without previous optimisation over 
local operations, see \cite{gerrtele}). 
This counterexample only arose for $\xi_1\neq\xi_2$. 
For $\xi_1=\xi_2$ 
the difference in teleportation fidelities mirrors faithfully the difference in logarithmic negativities. 

\subsection*{\bf \normalsize \textsf{Summary}}

We have considered continuous variable quantum memories operated in atomic ensembles by QND feedback, and 
meaningfully compared the storage of single-mode squeezing with the storage of two-mode entanglement to find that:
\begin{itemize}
\item{if the predominant additive noise in each memory acts on a single quadrature 
(ideal or near-ideal memories) and for comparable loss factors in the two memories, 
then the storage preserving the most entanglement is the 
one whereby the quadrature subject to the largest noise is made more robust 
(larger);}
\item{if the differences $\delta_{q}$ and $\delta_p$ 
in the noises on the $\hat{q}$ and $\hat{p}$ quadratures influence the final state in opposite ways, 
the advantage offered by the storage of squeezing proves to be more stable 
than the one granted by the storage of entanglement
({\em i.e.,} for $s>1$, $\delta_p\ge \delta_q\ge 0$ $\Rightarrow$ $\delta E_{\N}\ge 0$ 
but $\delta_q\ge \delta_p\ge 0$ $\nRightarrow$ $\delta E_{\N}\le 0$);}
\item{in the case of phase-insensitive Gaussian noise ($y_{qi}=y_{pi}$ for $i=1,2$), 
storing squeezing is advantageous;}
\item{the optimality of the storage of squeezing is stable under asymmetric variations of the loss 
factors between the two memories (the same is not true for the storage of entanglement).}
\end{itemize}
These findings were confirmed 
in terms of both entanglement -- as quantified by the logarithmic negativity -- 
and teleportation efficiency of the respective final states.

Notice that we did not consider the {\em optimised} production or protection of entanglement over 
any passive operation as in \cite{passive} and \cite{optim} 
(which would require an accurate knowledge of the input state 
and of the action of the memory), since we 
rather tried to identify conditions that would hold for 
generic inputs, and that would point to the 
best choice for a retrieval of the entanglement on demand. 

Our results could have significant impact as operational guidelines 
for the storage and retrieval of 
continuous variable entanglement, which will be a ubiquitous prerequisite 
in the areas of quantum communication and information processing alike.

\subsection*{\normalsize \bf \textsf{Appendix -- Derivation of analytical results}}
Although the smallest symplectic eigenvalues $\tilde{\nu}_a$ and $\tilde{\nu}_b$ could 
in principle be evaluated analytically, their expressions are extremely complex, and do not 
allow for direct manipulation.
To treat our comparison analytically, 
we will instead first assume $\xi_1=\xi_2=1$ and
define a class of covariance matrices $\sig(\vartheta)$:
$
\sig_\vartheta = R \sig_0 R^{\sf T} + R_{\vartheta} Y  R_{\vartheta}^{\sf T}
$, 
with $R_{\vartheta}={\rm e}^{\vartheta J}$, and $J$ with entries 
$J_{jk}=\delta_{j+2,k}-\delta_{j,k+2}$, 
such that $R_{\pi/4}=R$ and, hence, $\sig_0=\sig_a$ and $\sig_{\pi/4}=\sig_b$.
The infinitesimal variations of the symplectic invariants with respect to $\vartheta$ 
can be obtained 
by applying the formula 
$\det(M+\vartheta N) = \det M + \det(N) \tr(MN^{-1}) \vartheta + O(\vartheta^2)$,
whence one gets:
\be
\frac{{\rm d}\tilde{\Delta}(\sig_\vartheta)}{{\rm d}\vartheta} = 
\left[(a-c)\delta p + (b-d)\delta q -4 \delta p\delta q \sin(2\vartheta) \right] \cos(2\vartheta) \, , \label{ddelta} 
\ee
\be
\frac{{\rm d}\det\sig_\vartheta}{{\rm d}\vartheta} = 
[ A\delta p + 
B\delta q +  
\delta p \delta q - 
 \delta p \delta q \sin(2\vartheta)] \cos(2\vartheta) \, , \label{ddet}
 \ee
 with 
$A=(b-d)(a+y_{q2})(c+y_{q2})$, 
$B=(a-c) (b+y_{p2})(d+y_{p2})$, 
$C=(cd-ab+(c-a)y_{p2}+(d-b)y_{q2})$ and 
$D=(a-c) (b-d)$,
$\delta p = y_{p2}-y_{p1}$ and $\delta q = y_{q2} - y_{q1}$ and 
$a=N_1 s$, $b=N_1/s$, $c=N_2/s$, $d=N_2 s$. 
\eq{numin} implies 
\be
\sqrt{\tilde{\Delta}^2(\sig_\vartheta)-4\det\sig_\vartheta}\frac{{\rm d}\tilde{\nu}_{\vartheta}^2}{{\rm d}\vartheta} = 
\frac{{\rm d}\det\sig_\vartheta}{{\rm d}\vartheta} - 
\tilde{\nu}_{\vartheta}^2 \frac{{\rm d}\tilde{\Delta}(\sig_\vartheta)}{{\rm d}\vartheta} \; . \label{dnu}
\ee

For $\delta p = y_{p2}=0$, which is the case for ideal memories, 
inserting (\ref{ddelta}) and (\ref{ddet}) into (\ref{dnu}) yields
\be
\frac{{\rm d}\tilde{\nu}_{\vartheta}^2}{{\rm d}\vartheta} = 
\frac{\left[bd(a-c)\delta q - \tilde{\nu}_{\vartheta}^2 (b-d) \delta q\right]}
{\sqrt{\tilde{\Delta}^2(\sig_\vartheta)-4\det\sig_\vartheta}}\cos(2\vartheta) \; .
\ee
Now, the condition (\ref{assume}) implies 
$(a-c)=(N_1s-N_2/s)\ge 0$ and $(b-d)=(N_1/s-N_2 s)\le 0$, so that 
${{\rm d}\tilde{\nu}_{\vartheta}^2}/{{\rm d}\vartheta}=k \delta q$,
where $k$ is positive for $0\le\vartheta\le \pi/4$.
Therefore, $\tilde{\nu}_a=\tilde{\nu}_0 \le \tilde{\nu}_{\pi/4}=\tilde{\nu}_{b}$ if 
$\delta q\ge 0$ and $\tilde{\nu}_{a}=\tilde{\nu}_0 \ge \tilde{\nu}_{\pi/4}=\tilde{\nu}_{b}$ if 
$\delta q\le 0$, which leads to (\ref{finda}) because of \eq{logneg}. 

If $\xi_1=\xi_2\neq1$, {\em i.e.\ } if the losses of the two memory cells are equal, 
one has $[X,R]=0$, such that Eq.~(\ref{siga}) becomes 
$\sig_a = R X \sig_0 X^{\sf T}R{\sf T}+Y$ and one can redefine the initial CM as 
$\sig_0'=X\sig_0 X^{\sf T}$, which takes the same form as $\sig_0$ by redefining 
$N_i\rightarrow N_i \xi_i^2$ for $i=1,2$. 
Our previous proof of the criterion $y_{q2}\ge y_{q1} \Leftrightarrow E_{\N}(\varrho_{a}) \ge E_{\N}(\varrho_b)$ 
then applies under the same condition (\ref{assume}), which is unaffected by the 
transformation of the $N_i$'s since $\xi_1=\xi_2$.  \proofend

\noindent Notice that $\sigma_0'$ might not be a physical CM, in that it could violate Robertson-Schr\"odinger 
inequalities. This is however irrelevant for our proof.\bigskip
 
\noindent {\bf \textsf{Acknowledgments.}} We thank K.~Jensen, T.~Fernholz and E.~S.~Polzik for a prompt 
clarification concerning the Niels Bohr Institute's QND feedback 
quantum memories experimental setting.


\begin{thebibliography}{99}

\bibitem{review}{C. Simon, M. Afzelius, J. Appel, A. Boyer de la Giroday, S.J. Dewhurst, N. Gisin, C.Y. Hu, F. Jelezko, S. Kroll, J.H. Muller, J. Nunn, E. Polzik, J. Rarity, H. de Riedmatten, W. Rosenfeld, A.J. Shields, N. Skold, R.M. Stevenson, R. Thew, I. Walmsley, M. Weber, H. Weinfurter, J. Wrachtrup, and R.J. Young, 
Eur. Phys. J. D {\bf 58}, 1 (2010).}
\bibitem{cvrmp}{Braunstein van Loock, Rev. Mod. Phys. {\bf 77}, 513 (2005).}
\bibitem{cvcrypto}{L. P. Lamoureux, E. Brainis, D. Amans, J. Barrett, and S. Massar, 
Phys. Rev. Lett. {\bf 94}, 050503 (2005); A. Leverrier and Ph. Grangier, 
Phys. Rev. Lett. {\bf 102}, 180504 (2009).}
\bibitem{passive}M. M. Wolf, J. Eisert, and M. B. Plenio, Phys. Rev. Lett. 90, 047904 (2003).
\bibitem{eit1}{J. Appel, E. Figueroa, D. Korystov, M. Lobino, and A. I. Lvovsky, Phys. Rev. Lett. {\bf 100}, 093602 (2008).}
\bibitem{eit2}{K. Honda, D. Akamatsu, M. Arikawa, Y. Yokoi, K. Akiba, S. Nagatsuka, T. Tanimura, A. Furusawa, and M. Kozuma, Phys. Rev. Lett. {\bf 100}, 093601 (2008).}
\bibitem{rmp}{K.~Hammerer, A.~S.~S{\o}rensen, and E.~S.~Polzik, Rev. Mod. Phys. {\bf 82}, 1041 (2010).}
\bibitem{expmemo}{B. Julsgaard, J. F. Sherson, J. Fiurasek, J. I. Cirac, and E. S. Polzik, Nature (London) {\bf 432}, 482 (2004).}
\bibitem{exptele}{J. F. Sherson, H. Krauter, R. K. Olsson, B. Julsgaard, K. Hammerer, J. I. Cirac, and 
E. S. Polzik, Nature (London) {\bf 443}, 557 (2006).}
\bibitem{squeezememo}{K. Jensen, W. Wasilewski, H. Krauter, T. Fernholz, B. M. Nielsen, A. Serafini, M. Owari, M. B. Plenio, M. M. Wolf, and E. S. Polzik, arXiv:1002.1920, to appear on Nature Physics (2010).}
\bibitem{jensin}M. B. Plenio and J. Eisert, Int. J. Quant. Inf. {\bf 1}, 479 (2003).
\bibitem{Gchannels}B. Demoen, P. Vanheuverzwijn, A. Verbeure, Lett. Math. Phys. {\bf 2}, 161 (1977); 
J. Eisert and M. M. Wolf, in {\em Quantum Information with Continous Variables of Atoms and Light}, N. J. Cerf, G. Leuchs, and E. S. Polzik Eds. (Imperial College Press, London, 2007).
\bibitem{nega}J. Lee, M. S. Kim, Y. J. Park, and S. Lee, J. Mod. Opt. {\bf 47}, 2151 (2000); 
J. Eisert, PhD Thesis (University of Potsdam, 2001);  
G. Vidal and R. F. Werner, Phys. Rev. A {\bf 65}, 032314 (2002);  
M. B. Plenio, Phys. Rev. Lett. {\bf 95}, 090503 (2005).  
\bibitem{serafozzi}{A. Serafini, F. Illuminati, and S. De Siena, J. Phys. B {\bf 37}, L21 (2004); 
A. Serafini, Phys. Rev. Lett. {\bf 96}, 110402 (2006).}
\bibitem{piro}{S.~Pirandola and S.~Mancini, Laser Physics {\bf 16}, 1418 (2006).}
\bibitem{gerrtele}{G.~Adesso and F.~Illuminati, Phys. Rev. Lett. {\bf 95}, 150503 (2005).}
\bibitem{optim}{N. Schuch, M. M. Wolf, and J. I. Cirac, Phys. Rev. Lett. {\bf 96}, 023004 (2006).}
\end{thebibliography}
\end{document}